\documentclass[10pt,conference]{IEEEtran}
\IEEEoverridecommandlockouts

\usepackage{subfig}

\usepackage [english]{babel}
\usepackage [autostyle, english = american]{csquotes}
\MakeOuterQuote{"}

\usepackage{printlen}
\usepackage{booktabs} 
\usepackage{tablefootnote}
\usepackage{footmisc} 
\usepackage{cite}
\usepackage{amsmath}
\usepackage{graphicx}

\begin{document}

\title{Enabling Ultra-Dense, Open-RAN, Vehicular Networks with Non-Linear MIMO Processing}

\author{\IEEEauthorblockN{George N. Katsaros and Konstantinos Nikitopoulos}\\ \vspace{-12pt}
 \IEEEauthorblockA{Wireless Systems Lab, 5G and 6G Innovation Centre, Institute for Communication Systems \\
 University of Surrey, Guildford, GU2 7XH, UK \\
 {\{g.katsaros,k.nikitopoulos\}}@surrey.ac.uk}
}

\maketitle

\begin{abstract}
Future autonomous transportation systems necessitate network infrastructure capable of accommodating massive vehicular connectivity, despite the scarce availability of frequency resources.
Current approaches for achieving such required high spectral efficiency, rely on the utilization of Multiple-Input, Multiple-Output (MIMO) technology.
However, conventional MIMO processing approaches, based on linear processing principles, leave much of the system's capacity heavily unexploited. They typically require a large number of power-consuming antennas and RF-chains to support a substantially smaller number of concurrently connected devices, even when the devices are transmitting at low rates. This translates to inflated operational costs that become substantial, particularly in ultra-dense, metropolitan-scale deployments. 
Therefore, the question is how to efficiently harness this unexploited MIMO capacity and fully leverage the available RF infrastructure to maximize device connectivity.
Addressing this challenge, this work proposes an Open Radio Access Network (Open-RAN) deployment, with Massively Parallelizable Non-linear (MPNL) MIMO processing for densely deployed, and power-efficient vehicular networks.
For the first time, we quantify the substantial gains of MPNL in achieving massive vehicular connectivity with significantly reduced utilized antennas, compared to conventional linear approaches, and without any throughput loss.
We find that an Open-RAN-based realization exploiting the MPNL advancements can yield an increase of over 300\% in terms of concurrently transmitting single-antenna vehicles in urban mobility settings and for various Vehicle-to-Infrastructure (V2I) and Network (V2N) use cases. In this context, we discuss how implementing MPNL within the Open-RAN ecosystem allows for simpler and more densely deployed radio units, paving the way for fully autonomous and sustainable transportation systems.

\end{abstract}


\IEEEpeerreviewmaketitle

\section{Introduction}\label{s_intro}

The transition to more intelligent, transportation systems promises to improve road safety, traffic efficiency, and user convenience by allowing seamless communication between vehicles, infrastructure, pedestrians, and network services, laying the foundation for fully autonomous driving \cite{bishen_vehicular_2021}. 
However, this transition necessitates a large number of vehicles, infrastructure, and pedestrians to be continuously connected, receiving and transmitting sensory information.
For example, a single busy intersection could accommodate over 400 concurrently transmitting vehicles and four times as many passengers and pedestrians \cite{karunathilake_survey_2022}. 
With the Federal Communications Commission (FCC) currently allocating 30 MHz of bandwidth for initial Cellular Vehicle-to-Everything (C-V2X) deployments, it becomes crucial to realize highly spectral-efficient solutions capable of supporting a massive number of connected devices and vehicles over the same frequency resources.

In the latest mobile generations, advancements in Multiple-Input Multiple-Output (MIMO) and Massive MIMO (mMIMO) technologies have been the primary drivers behind device connectivity and throughput gains. Particularly in the sub-6 GHz frequency range (FR1), where the multipath propagation of metropolitan environments enables substantial spatial multiplexing gains \cite{goldsmith_capacity_2003}.
Still, the existing, conventional MIMO approaches, (e.g., based on linear Zero Forcing (ZF) and Minimum Mean Square Error (MMSE) principles) leave much of the system's capacity unexploited and rely on the employment of a substantial number of antennas and RF chains at the access points, to serve a much smaller number of concurrently transmitted user streams. 
In fact, as demonstrated later in this work, linear approaches might require even three times the necessary number of antennas, to serve vehicles and devices that may even be transmitting at low rates.
This raises substantial concerns regarding the operational costs
and the carbon footprint of such "inflated" radio infrastructure \cite{holtkamp2013minimizing,auer2011much}, especially in densely-deployed metropolitan environments. Therefore the current challenge is to design novel processing approaches, capable of practically delivering the MIMO capacity gains, maximizing device connectivity without compromising the power efficiency of the radio access network (RAN).

Open-RAN \cite{azariah2022survey} with its standardized open interfaces and its multiple disaggregation options, 
allows for a highly diversified RAN ecosystem, and high flexibility with programmatic control, essentially creating the conditions for catalyzing innovation \cite{noauthor_open_2022}. Moreover, Open-RAN, through its proposed functional splits \cite{ORANARCH2023} appears instrumental in streamlining the radio/roadside unit (RU) architecture by shifting resource-hungry processes of the physical layer (PHY)—or even the entire PHY—from the radio end to more centralized locations. This allows, apart from a simplification of the radio unit architecture, the flexible employment and rapid integration of more sophisticated MIMO-PHY processing schemes, such as nonlinear (NL) MIMO processing. As we demonstrate in this work, NL processing, has the potential to significantly enhance vehicular connectivity capabilities, supporting substantially more concurrently transmitting vehicles over very limited frequency resources and without further increasing the number of access point antennas.

\begin{figure}[t]
    \centering
    \includegraphics{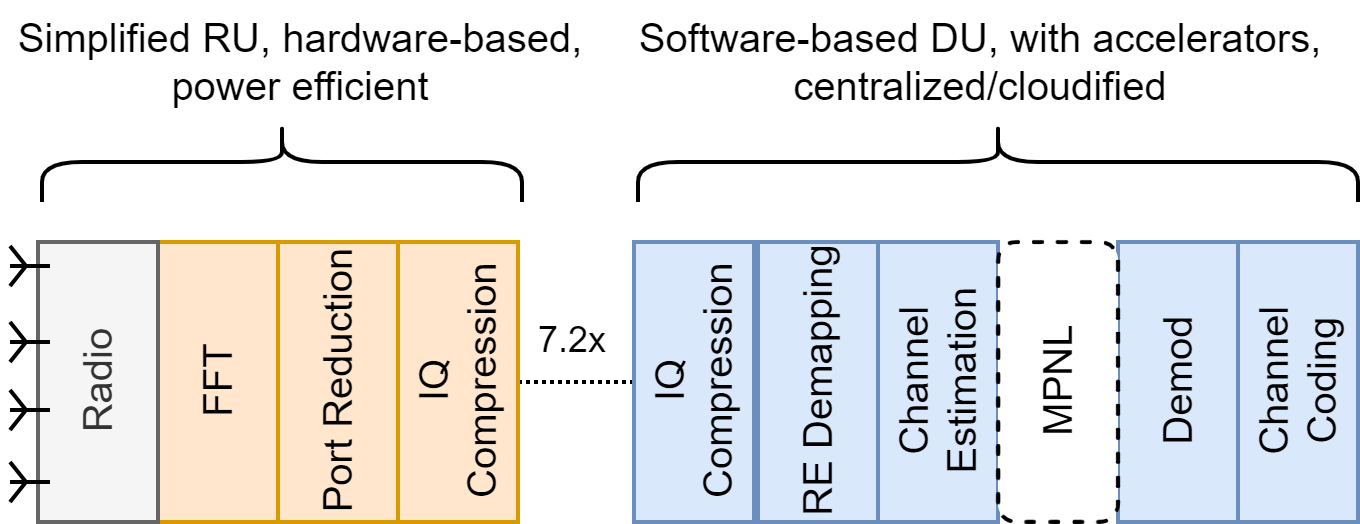}
    \caption{Open-RAN 7.2x functional split, for ultra-dense, urban V2I/V2N deployments with MPNL processing}
    \vspace{-12pt}
    \label{fig:oranmpnl}
\end{figure}

Still, the processing latency of conventional nonlinear MIMO processing methods (for instance, those based on the Sphere Decoder \cite{burg_vlsi_2005,nikitopoulos_geosphere_2014})  renders them unsuitable for real-time MIMO implementations. That is because of their exponentially scaling computational complexity with the number of concurrently supported MIMO streams \cite{jalden_complexity_2005,hassibi_sphere-decoding_2005} and their weak parallelization properties. To counteract this, recent proposals like massively parallelizable nonlinear (MPNL) processing \cite{nikitopoulos_massively_2022,husmann_flexcore_2017,nikit_worldpat} have been introduced, aiming to alleviate the processing latency concerns by enabling the aggressive parallelization of related functionalities and exploiting, fully, the capabilities of state-of-the-art commercial off-the-shelf (COTS) parallel processing platforms (eg., CPUs, GPUs, FPGAs). Despite the potential of MPNL processing, a practical realization of MPNL has not been discussed in the context of Open-RAN, or within the highly-dynamic environment of vehicular communications. Consequently, the true system-level connectivity gains offered by MPNL are yet to be fully understood, along with their implications, for example, in system-level power consumption, within a real-world, ultra-dense V2X setting.

In this work, we propose and evaluate MPNL MIMO processing as an enabler of ultra-dense Open-Ran vehicular networks.
For the first time, we quantify the gains of MPNL in terms of concurrently transmitting vehicles in a metropolitan setting with very limited available spectrum resources. We present, for different Vehicle-to-Infrastructure (V2I) and Network (V2N) use cases \cite{bishen_vehicular_2021}, and various numbers of employed RU antennas, an average increase of over 300\% in the number of concurrently transmitting vehicles compared to the conventional linear approaches (i.e., MMSE).
Moreover, we find that MPNL can achieve a reduction of more than 68\% in the total number of employed antennas without any packet error rate (PER) performance loss.
Interestingly, we showcase scenarios where MPNL can support a much larger number of MIMO streams than base-station antennas, even up to three times as many, and without the need for any traditional Non-Orthogonal Multiple Access (NOMA) \cite{6692652} schemes.
Finally, we discuss how those gains, when realized in an Open-RAN context, can be directly translated into power savings, practically enabling future ultra-dense deployments of roadside C-V2X infrastructure.

\section{Background}\label{s_adopted}

Traditional linear-based MIMO algorithms tackle the detection problem by converting the MIMO channel into separate SISO (Single-Input Single-Output) channels. Such transformation simplifies the processing demands and allows the application of well-established SISO-based techniques to MIMO systems. However, due to this linear transformation, a significant portion of the MIMO capacity remains underutilized, constraining the achievable throughput and the number of streams that can be concurrently transmitted.
On the other hand,  traditional non-linear approaches \cite{jalden_complexity_2005,hassibi_sphere-decoding_2005} focus on jointly processing the mutually interfering MIMO streams, accounting for the MIMO channel characteristics, without degrading its inherent capabilities. Even though traditional non-linear processing approaches promise substantial throughput and connectivity gains, their exponentially scaling computational complexity with the number of concurrently supported MIMO streams \cite{jalden_complexity_2005}
and their weak parallelization properties often render them unsuitable for practical, real-time realizations.

In response to those challenges, the Massively Parallelizable Non-Linear (MPNL) Processing framework was recently introduced \cite{nikitopoulos_geosphere_2014,husmann_flexcore_2017,nikitopoulos_massively_2022}, that targets to deliver the gains of non-linear processing into a practically implementable solution. 
At its principle, MPNL has the ability to identify the relative position to any possible transmission vector of the most promising solutions to the non-linear transmission problem by simply observing the MIMO channel before any transmission begins. Then, during transmission, the actual most promising solutions ($N_p$) are identified and processed in parallel in order to solve the non-linear transmission problem. 
Identifying and processing only the most promising solutions to the non-linear problem substantially decreases the required computational complexity while preserving the achievable performance. In addition, the intrinsic parallelism of MPNL facilitates its efficient integration with modern, highly parallel computing platforms.
While earlier works and implementations \cite{husmann_flexcore_2017} present some first evaluations of the throughput gains of MPNL approaches, this study is the first to systematically examine the substantial connectivity gains of Open-RAN-enabled MPNL, particularly within the highly dynamic environment of vehicular communications.

\begin{figure}[t]
    \centering
    \includegraphics[height=4cm]{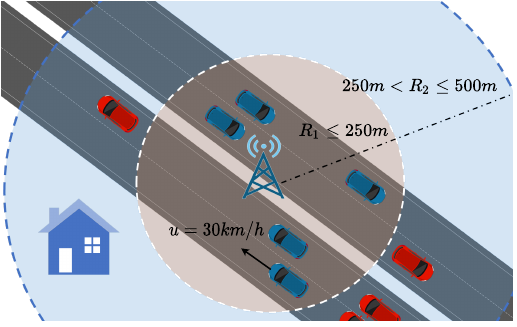}
      \caption{The simulation setup, utilizing CDL-D channel models, examining two coverage regions ($R_1$ within 250 m from the RU and $R_2$ within 250-500 m) with  vehicle speed set to 30 km/h.}
      \vspace{-10pt}
    \label{fig:simsetup}
\end{figure}

\subsection{Functional Split Deployment}

The introduction of multiple functional splits \cite{O-RAN2023arch} between the Radio Unit (RU) and the Distributed Unit (DU), expands deployment flexibility, allowing for a range of MPNL deployments, each tailored for specific communication scenarios. Our examined Open-RAN split option is the 7.2x split option as presented in Fig.\ref{fig:oranmpnl}. In 7.2x, the core PHY processing functions residing on the RU are the Fourier transforms, while the rest of the PHY processing is implemented on DU. This split-based deployment option is best suited for urban and denser RU deployments. It offers a balance between the RU design complexity and the fronthaul bandwidth requirements. The simplification of the RU processing can result in reductions in its computational power consumption and can unlock the ability to centralize DU processing by serving multiple RUs with the same DU infrastructure. Moreover, since the computationally intensive processing can be moved away from the radios, new deployment options for PHY are being unlocked. Such deployments include heavily software-based processing, which has the potential to be realized even on existing cloud service infrastructure.
This allows novel algorithmic approaches such as MPNL to harness flexibly, parallel processing resources (often heterogeneous), elastically scale based on the RU traffic conditions, and allocate their computing power on the radio units where it is most required (e.g., the ones serving higher traffic at a given time). As we discuss later in Section \ref{wayforward}, such deployment flexibility has the potential to deliver substantial power gains on a system level.

\begin{figure}[t]
    \centering
    \includegraphics{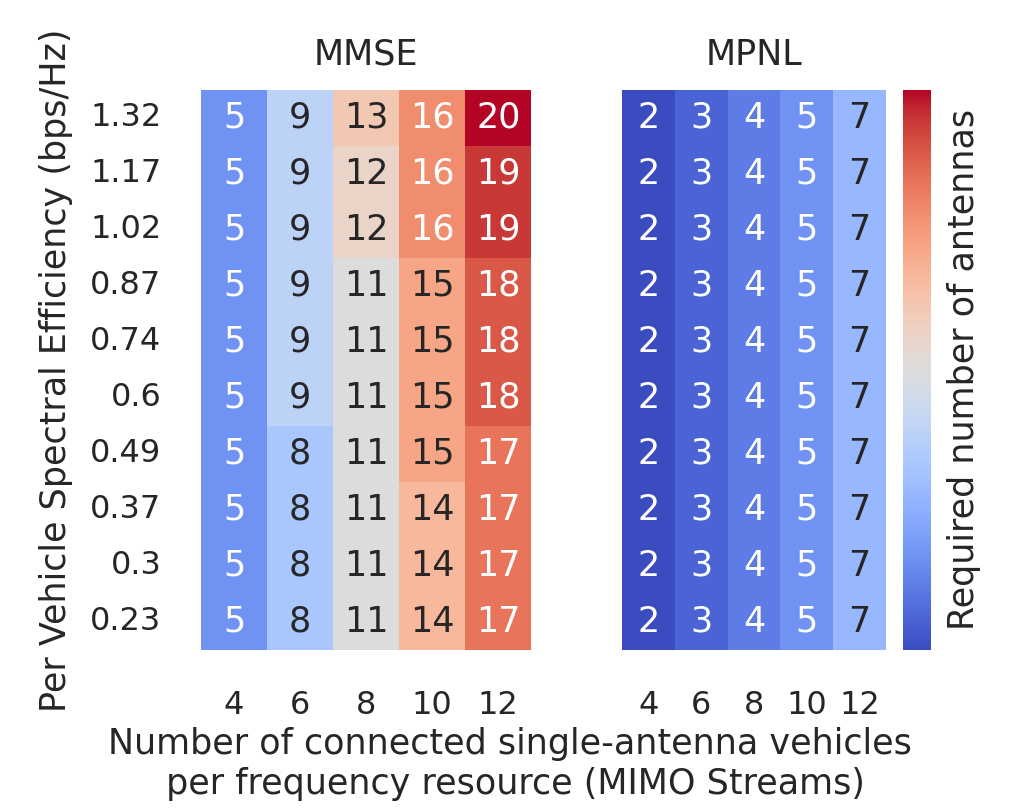}
    \caption{The number of required antennas for supporting 10 percent PER vehicles (30km/h, CDL-B) for different rates and MIMO orders, when using MMSE (left), MPNL (right).}
    \label{fig:heatmap}
    \vspace{-10pt}
\end{figure}

\begin{figure*}[t]
\centering
    \includegraphics[height=6cm]{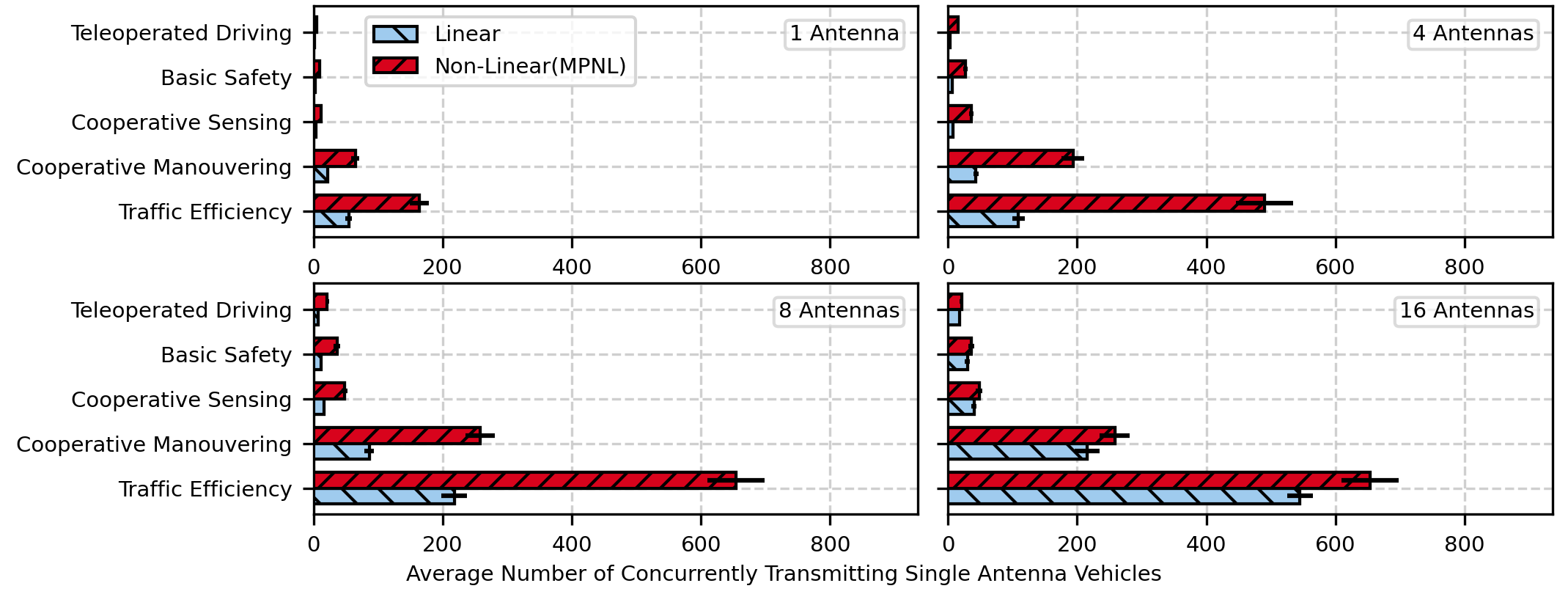}
\caption{Comparison of the average number of concurrently supported single antenna vehicles for different RU sizes, utilizing linear (MMSE) and non-linear processing. The results are for indicative V2I/V2N use cases assuming urban vehicular mobility at 30 km/h within $R_2$ distance (i.e., 250–500 m Fig.\ref{fig:simsetup}) from the RU with average Rx SNR at 15 dB $\pm1$dB, and a maximum of 12 supported MIMO streams.}
\vspace{-10pt}

\label{f_lowrate_conn}
\end{figure*}
\section{Simulation Methodology} \label{s_adopted_sim}

\subsection{Channel Generation}
In assessing MPNL's connectivity gains in the context of Open-RAN-based vehicular networks,
we examined various MU-MIMO scenarios with different MIMO dimensions.
For each MIMO dimension evaluated, we used a unique set of 50 independent Cluster Delay Line (CDL) \cite{3GPPTR38901} MIMO channels.
We specifically analyzed MU-MIMO systems that support up to 32 antennas at the radio unit, and  2, 4, 8, and 12 MIMO streams, corresponding to individual single-antenna vehicles transmitting on the same frequency resource. 

Our evaluations, presented in Section \ref{s_discussion}, explore a dense urban mobility scenario, with vehicles at a constant speed of 30km/h.
We utilized CDL-D channels, which simulate an outdoor setting and include both line of sight (LOS) and non-LOS components. For the channel generation, the vehicles were randomly distributed, within the two regions, as shown in Fig.\ref{fig:simsetup}. We also assumed that the relative vehicle movement is either towards or away from the radio unit and that the communication is being established at the 3.5GHz carrier frequency. Within each of the examined regions (i.e., R1 and R2), it is assumed that the path loss is almost uniform. The channels were generated utilizing MATLAB's 5G Toolbox \textit{nrCDLChannel}  
\begin{table}[h]
\centering
\begin{tabular}{lc}
\toprule
\textbf{Use Case} & \textbf{Data Rate Requirement (Mbps)} \\
\midrule
Teleoperated Driving & 50 \\
Basic Safety & 30 \\
Cooperative Sensing & 25 \\
Cooperative Manouvering & 5 \\
Traffic Efficiency & 2 \\
\bottomrule
\end{tabular}
\caption{Data rate requirements for different indicative V2I/V2N use cases.}
\label{tab:usecase}
\end{table}

\subsection{Comparative Simulations}
To make comparative assessments, we conducted uplink simulations utilizing both the MPNL and the de facto standard linear MMSE detection approach. 
The generated channels were grouped based on the number of supported MIMO streams $N\in\{4, 6, 8, 10, 12\}$. Each group comprised of 50 channels for each number of employed base station antennas $M$, where $M\in[2, 32]$. For each examined number of MIMO streams, starting
from the smallest number of employed antennas, we evaluated for all the corresponding
channels the average PER performance of each examined detector. If the resulting PER didn’t
satisfy our 10\% maximum threshold, we increased the number of base station antennas by one
and repeated the experiment either until the performance reached the acceptable threshold or
until the number of antennas reached the maximum examined in our simulations, i.e., 32; in such
case, the examined scenario is considered not supported

The average receiver SNR is set at 15$\pm1$dB within R1 and 20$\pm1$dB within R2 (Fig. \ref{fig:simsetup}.)
The packet error rate (PER) performance was evaluated after the Low-Density Parity-Check (LDPC) decoding stage, with a maximum allowance of 10 LDPC iterations.
For the non-linear processing, we assumed a maximum of 32 parallel paths ($N_p$).
Furthermore, we assumed a transmission bandwidth of 30MHz within the 5G-NR's Frequency Range 1 (FR1), at the subcarrier spacing of 30kHz, resulting in a slot duration of half a millisecond. Each slot is presumed to consist of 12 data symbols and 2 symbols reserved for the demodulation reference signal (DMRS). We focussed on uplink heavy slots because, in the context of vehicular communications, they can correspond to a lower bound on the presented connectivity gains. This is because, unlike traditional communications, V2X traffic is envisioned to be heavily uplink-based \cite{yang_can_2019}, requiring substantial data volumes, from multiple cameras and on-board sensors to be transmitted by the vehicles to the network infrastructure. While not in the scope of this study, it is worth mentioning the MPNL framework incorporates non-linear precoding in the downlink by enabling practical Vector Perturbation \cite{VP,improvedvp,gain15dbvp,FCSE} with similar gains as in the uplink, compared to linear precoding.

\section{Results and Discussion}\label{s_results}

This section discusses the potential vehicular connectivity gains of MPNL.
Figure \ref{fig:heatmap} presents the gains of our proposed MPNL MIMO processing compared to MMSE processing in terms of the number of active RU antennas required for achieving a 10\% PER for different per-vehicle rates and for different numbers of concurrently transmitting vehicles (i.e., up to 12 MIMO streams) randomly placed within R2 (Fig.\ref{fig:simsetup}). 
As shown, MPNL's benefits in reducing the required number of antennas without PER loss compared to MMSE are substantial. 
In all the examined cases, the number of concurrently supported MIMO streams exceeds the number of employed antennas, with corresponding gains of nearly 70\% fewer RU antennas than the equivalent MMSE-based deployment. The highest gains appear when supporting 12 MIMO concurrent streams, where MPNL requires just 7 employed antennas while MMSE would require almost three times as many to achieve the same PER performance of 10\%. 
MPNL's ability to support, for the same number of antennas, a much larger number of MIMO streams, paves the way for ultra-dense deployment scenarios, facilitating massive connectivity use cases.
It is worth mentioning that MPNL gains are more resilient to code-rate changes within the same constellation group (i.e., QPSK for MCS less than or equal to 9). In contrast, utilizing traditional linear MMSE, increasing the rate often suggests increasing the minimum required antennas even if the transmitted constellation remains the same.
Lastly, while an increase in the vehicular speed within the range of 30-70 km/h (typical urban mobility speed) presents an observable increase in the PER, it does not recommend employing additional antennas, as it does not surpass the targetted 10\% PER within the examined speed range.
In the following paragraphs, we attempt to quantify and translate those gains into practical connectivity gains for specific V2I/V2N use cases.

\begin{figure*}[ht]
\centering
    \includegraphics[height=6cm]{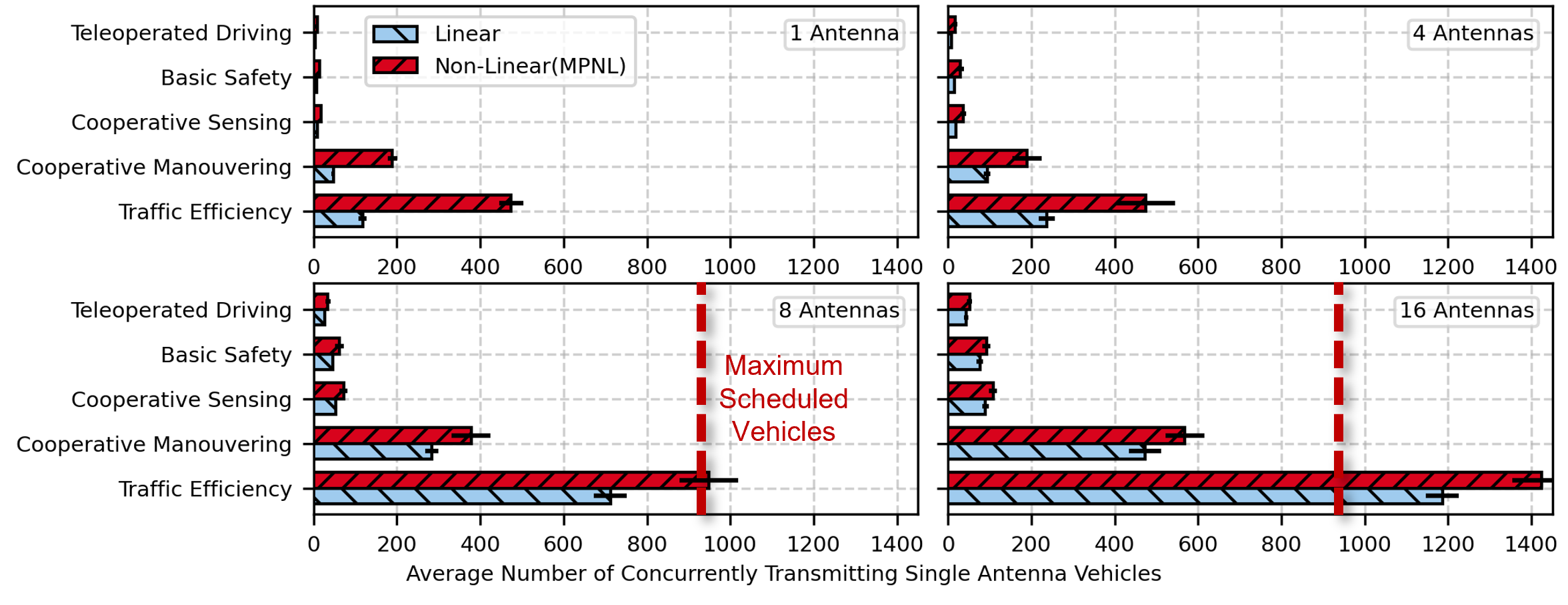}
\caption{Comparison of the average number of concurrently supported single antenna vehicles for different RU sizes, utilizing linear (MMSE) and non-linear processing. The results are for indicative V2I/V2N use cases assuming urban vehicular mobility at 30 km/h within $R_1$ (i.e., $\leq 250m$ Fig.\ref{fig:simsetup}) from the RU with average Rx SNR at 20 dB $\pm1$ dB, and a maximum of 12 supported MIMO streams.}    
\vspace{-10pt}
\label{fig.high_rate_conn}
\end{figure*}

To investigate how the MPNL's antenna gains can translate into practical connectivity gains offered by non-linear processing, we've examined five indicative V2I/V2N use cases, each presenting a different uplink data rate requirement \cite{garcia_tutorial_2021,bishen_vehicular_2021,xiong2022reducing}.
The examined scenarios and their required data rate are summarized in Table \ref{tab:usecase}. The first examined use-case focuses on remote or Tele-operated Driving, for which we estimate a required uplink throughput of 50 Mbps.
The second is Basic Safety features, which comprise features such as collision avoidance, pedestrian safety, road condition alerts, and emergency vehicle alerts and are estimated to necessitate an uplink data rate of 30 Mbps. Third, we examine Cooperative Sensing at 25Mbps and cooperative maneuvering and traffic efficiency at 5 and 2 Mbps, respectively.

We separately examined four different radio units performing MMSE and MPNL MIMO detection with a fixed number of antennas: 1, 4, 8, 16, and a maximum of 12 single-antenna vehicles per scheduling unit (MIMO streams).
We utilized a scheduling unit of 1 resource block, thereby, according to 5G-NR numerology for 30 MHz of transmission bandwidth and 30 KHz for the subcarrier spacing (SCS), a maximum of 
$78 \times N = 936$
concurrently transmitting vehicles can be scheduled within a slot, where $\mathrm{N=12}$ is the assumed maximum number of supported MIMO streams by our base station.
Based on the required data rate per vehicle for the given use cases and the estimated spectral efficiency for vehicles located within the two examined regions, we deduced the maximum number of vehicles that each use case can accommodate in a single slot. In general, the maximum number of supported vehicles per base station ($NV_{i}^{max}$) for a given use-case ($i$) can be estimated by Eq. (\ref{eq.maxveh}).
%

\begin{equation}
\mathrm{NV}_{i}^{\mathrm{max}} = \left\lfloor \frac{\mathrm{NRB}}{\left\lfloor \frac{R_i}{\mathrm{SE} \times \mathrm{SCS} \times \mathrm{SC}_{\mathrm{RB}}} \right\rfloor} \right\rfloor \times \mathrm{N}
\label{eq.maxveh}
\end{equation}

where ${R_i}$ is the uplink throughput requirement corresponding to the use case $i$, $\mathrm{SE}$ is the average per-vehicle spectral efficiency within the examined coverage area, $\mathrm{SCS}$ is the subcarrier spacing, $\mathrm{SC_{RB}}$ is the number of subcarriers per resource block and $\mathrm{NRB}$ is the total number of resource blocks in a symbol. 
Moreover, in cases where the throughput requirement of the examined application is sufficiently low compared to the achievable spectral efficiency of the vehicles, that is when $\frac{{{R}_{\text{{i}}}}}{{\mathrm{SE} \times \mathrm{SCS} \times \mathrm{SC}_{\mathrm{RB}}}} < 1 $, then the scheduling unit dictates the maximum supported vehicles in the slots $\mathrm{NV}_{i}^{\mathrm{max}} = NRB\times N$.
Given the relationship between the number of required base station antennas and supported MIMO streams for the linear and the nonlinear processing, as we previously discussed, we estimate the average number of vehicles that a fixed number of antennas can support under the 10\% PER constraint and for each given use-case.
Our results for urban mobility vehicles within $R_2$ ($250-500$ m) are presented in Fig. \ref{f_lowrate_conn}. Similarly, in Figure \ref{fig.high_rate_conn}, we present the connectivity gains of MPNL for vehicles supporting higher rates within $R_1$ ($<250$ m).

The gains of MPNL processing in terms of concurrently supported vehicles are, in most cases, between 200 and 400\%. The exception is at 16 base station antennas, where the gains are limited only by the maximum number of supported streams that we assumed for our simulations (i.e., 12 MIMO streams). In fact, non-linear processing can already support 12 MIMO streams with just 8 RU antennas (Fig.\ref{f_lowrate_conn}), which bring quantifiable power gains, as we discuss in Section \ref{wayforward}.
Moreover, in Fig. \ref{fig.high_rate_conn}, the maximum number of vehicles that can be scheduled for supporting the 2Mbps per vehicle required for Traffic Efficiency can be achieved with just 8 antennas and non-linear processing. 
It's important to acknowledge that signaling overhead can pose a practical constraint in achieving such high levels of vehicular connectivity. However, it pertains to the optimization of the control channels, a topic that exceeds the scope of our current study. However, it is noteworthy to mention strategies such as Semi-Persistent Scheduling (SPS), introduced in 5G-NR, which could alleviate such overhead. Instead of scheduling each individual transmission separately—which would necessitate a considerable amount of signaling — SPS enables the network to prearrange a series of transmissions at fixed intervals.

\subsection{Implications and the Way Forward} \label{wayforward}
As discussed in the previous section, supporting substantially more MIMO streams with the same number of RU antennas has the potential to boost vehicular connectivity in the order of 200–400\%.
Equivalently, MPNL gains can be harnessed by passively reducing the employed number of RU antennas while maintaining the same throughput and connectivity gains as traditional approaches. This has an immediate impact on the system-level power efficiency as it can be directly translated into power gains. For instance, assuming an indicative value of the power consumption of a single uplink RF chain at 15.6W \cite{gong_nonlinear_2021}, as previously presented in Fig. \ref{fig:heatmap}, MPNL can reduce the RU power consumption by nearly 70\%, which is equivalent to over 180W reduction for just a single access point.
Moreover, potential ways to further enhance the power efficiency of RUs is to enable a flexible and intelligent mechanism of activation and deactivation of the RF-Chains at high granularity, even on a per-symbol level, as it is being actively discussed within the Open-RAN community \cite{noauthor_open_nodate}. This can allow the matching of the corresponding RU consumption to the traffic requirements, maximizing MPNL gains.
Reducing RU consumption is particularly important for urban deployments as it allows for greater densification of the network infrastructure in metropolitan environments by reducing substantially not only the corresponding operating costs but also by paving the way towards more carbon-efficient and potentially even solar-powered RU infrastructure, eliminating deployment restrictions related to grid access.
From the DU perspective, the computational power costs of DU can be centralized and even cloudified. This can unlock even greater system-level power gains as the computational power can be flexibly allocated only to the required RUs. Similarly, the discussed passive power gains of MPNL can allow for a transition to fully software-based Open-RAN deployments \cite{softiphy,nikitopoulos2024towards} while maintaining the system-level energy efficiency levels of traditional hardware-based RANs. 
\label{s_discussion}

\section{Conclusions}
This work proposes an Open-RAN deployment of our Massively Parallelizable Non-Linear (MPNL) MIMO processing framework to enhance vehicular network connectivity with limited spectrum availability. For the first time, we quantify that MPNL can significantly outperform traditional linear processing (e.g., MMSE) by enabling over a 300\% increase in concurrent single-antenna vehicle transmissions across various V2I and V2N scenarios, while maintaining Radio Unit power output and per-vehicle throughput. Furthermore, we demonstrate that MPNL's reduction of required Radio Unit antennas by nearly 70\% boosts system-level power efficiency, thus facilitating the development of more sustainable, dense, and intelligent transportation systems.

\section*{Awknowledgements}
This work has been supported by UK’s Department for Science, Innovation, and Technology, “HiPer-RAN” project.

\bibliography{vehcom_lib}



\end{document}